%% file: main.tex
\documentclass[conference]{IEEEtran}
\usepackage{geometry}
\usepackage{cite}
\usepackage{amsmath,amssymb,amsfonts}
\usepackage{algorithmic}
\usepackage[acronyms]{glossaries}
\input{acronyms.tex}

\usepackage{graphicx}
\usepackage{textcomp}
\usepackage{xcolor}
\def\BibTeX{{\rm B\kern-.05em{\sc i\kern-.025em b}\kern-.08em
    T\kern-.1667em\lower.7ex\hbox{E}\kern-.125emX}}

\begin{document}
%\newgeometry{top=0.85in, bottom=1in, left=0.75in, right=0.75in}

% \title{Bridging Research and Standardization: Innovations and Methodology for 6G Standard Contributions}
% \title{Bridging Research and Standardization: \\ A Methodology for Integrating Innovations \\ into 6G Standard}
\vspace{-1.5em}
\author{\IEEEauthorblockN{Francesca Conserva\IEEEauthorrefmark{1},
Fabio Busacca\IEEEauthorrefmark{2},
Corrado Puligheddu\IEEEauthorrefmark{3},
Simone Bizzarri\IEEEauthorrefmark{4},
Maurizio Fodrini\IEEEauthorrefmark{4}, \\
Giampaolo Cuozzo\IEEEauthorrefmark{1}, and
Riccardo Marini\IEEEauthorrefmark{1}} \\
\vspace{-1.3em}

\IEEEauthorblockA{
\IEEEauthorrefmark{1} \small CNIT, National Laboratory of Wireless Communications (WiLab), Bologna, Italy; 
\IEEEauthorrefmark{2} \small Università di Catania, Italy; \\
\IEEEauthorrefmark{3} \small Politecnico di Torino, Italy; 
\IEEEauthorrefmark{4} \small Fibercop S.p.A., Italy
}

}
\vspace{-0.5em}

\title{Bridging Research and Standardization: Innovations and Methodology for 6G Standard Contributions
% \thanks{This work was partially supported by the European Union under the Italian National Recovery and Resilience Plan (NRRP) of NextGenerationEU, partnership on “Telecommunications of the Future” (PE0000001 - program “RESTART”).}
}

\IEEEaftertitletext{\vspace{-2.5\baselineskip}}
\maketitle
\vspace{-4\baselineskip}
% \vspace{-0.2cm}

\begin{abstract}
The transition towards 6G presents unique challenges and opportunities in mobile networks design and standardization. Addressing these challenges requires a robust methodology for analyzing and selecting innovations that can be effectively translated into \gls{3gpp} contributions. This paper presents a systematic approach to bridging research and standardization, ensuring that cutting-edge advancements extend beyond academia and translate into concrete standardization efforts. The proposed methodology has been applied within the Italian RESTART framework to two ongoing research areas: \glspl{mpn} and \glspl{ndt}, both key enablers of next-generation networks. \glspl{mpn} enhance dynamic adaptability and resource management, while \glspl{ndt} enable real-time simulation, predictive analytics, and intelligent decision-making. Their integration into 3GPP Release 20 will be instrumental in shaping a flexible and future-proof mobile ecosystem.
These innovations exemplify how research-driven solutions can align with 6G standardization objectives. By applying the proposed methodology, we aim to establish a systematic pathway for transitioning research into impactful \gls{3gpp} contributions, ultimately driving the evolution of next-generation networks.

% This paper introduces a structured methodology aimed at bridging research and standardization, ensuring that research outcomes on hot innovations non restano confinati all'accademia ma can be translated into concrete standardization contributions. This ensures that research extends beyond academia, facilitating a seamless transition from innovation to global adoption supporting the development of scalable and
% interoperable 6G systems.

% This methodology has been applied to research conducted
% within RESTART projects on MPNs and NDTs, highly rele-
% vant topics in the evolution of next-generation networks. MPNs
% enable dynamic adaptability and efficient resource manage-
% ment, while NDTs provide real-time simulation, predictive
% analytics, and intelligent decision-making
% Their integration
% into 6G Release 20 will be pivotal in shaping a flexible and
% future-proof mobile ecosystem. 

% Both innovations are the result of research conducted within the Italian RESTART framework, exemplifying how research-driven solutions can align with 6G standardization goals. By applying the proposed methodology, we aim to ensure a systematic transition from research results to impactful \gls{3gpp} contributions, ultimately supporting the development of flexible, efficient, and scalable 6G networks.

% highlighting \glspl{mpn} and \glspl{ndt} as key examples of innovative approaches to be considered for 6G. \glspl{mpn} enable cross-layer and cross-domain programmability, while \glspl{ndt} facilitate real-time simulation and optimization of mobile networks. 
\end{abstract}
% \vspace{-0.2cm}
\begin{IEEEkeywords}
6G, Standardization, 3GPP, Methodology, Morphable Programmable Networks, Network Digital Twin.
% \vspace{-0.3cm}
\end{IEEEkeywords}
\glsresetall
\section{Introduction}
\IEEEPARstart{T}{he} evolution towards 6G represents a critical step in advancing mobile communication systems, addressing the increasing demands of a hyper-connected and intelligent society \cite{6g_vision2023}. Future networks will be characterized by extreme data rates, ultra-low latency, pervasive connectivity, and enhanced energy efficiency, while integrating \gls{ai} as a fundamental enabler of automation, intelligence, and optimization \cite{BANAFAA2023,2024_6gpotential}. These advancements will unlock unprecedented adaptability, flexibility, and programmability, reshaping network design and operation to support emerging intelligent and reconfigurable network paradigms.

Leading this transformation, innovations such as \glspl{mpn}, \gls{ndt}, massive MIMO, mmWave communications, and Non-terrestrial networks (NTN) are driving breakthroughs across sectors such as healthcare, entertainment, and smart cities. These technologies foster a seamless convergence of the physical and digital worlds, enabling real-time, immersive, and highly interactive experiences \cite{BANAFAA2023}. Realizing such an ambitious vision requires a strong interplay between research and standardization. While research drives technological innovations, standardization ensures interoperability, scalability, and global adoption.
\begin{figure}[!t]
\centering
\includegraphics[width=0.9\columnwidth]{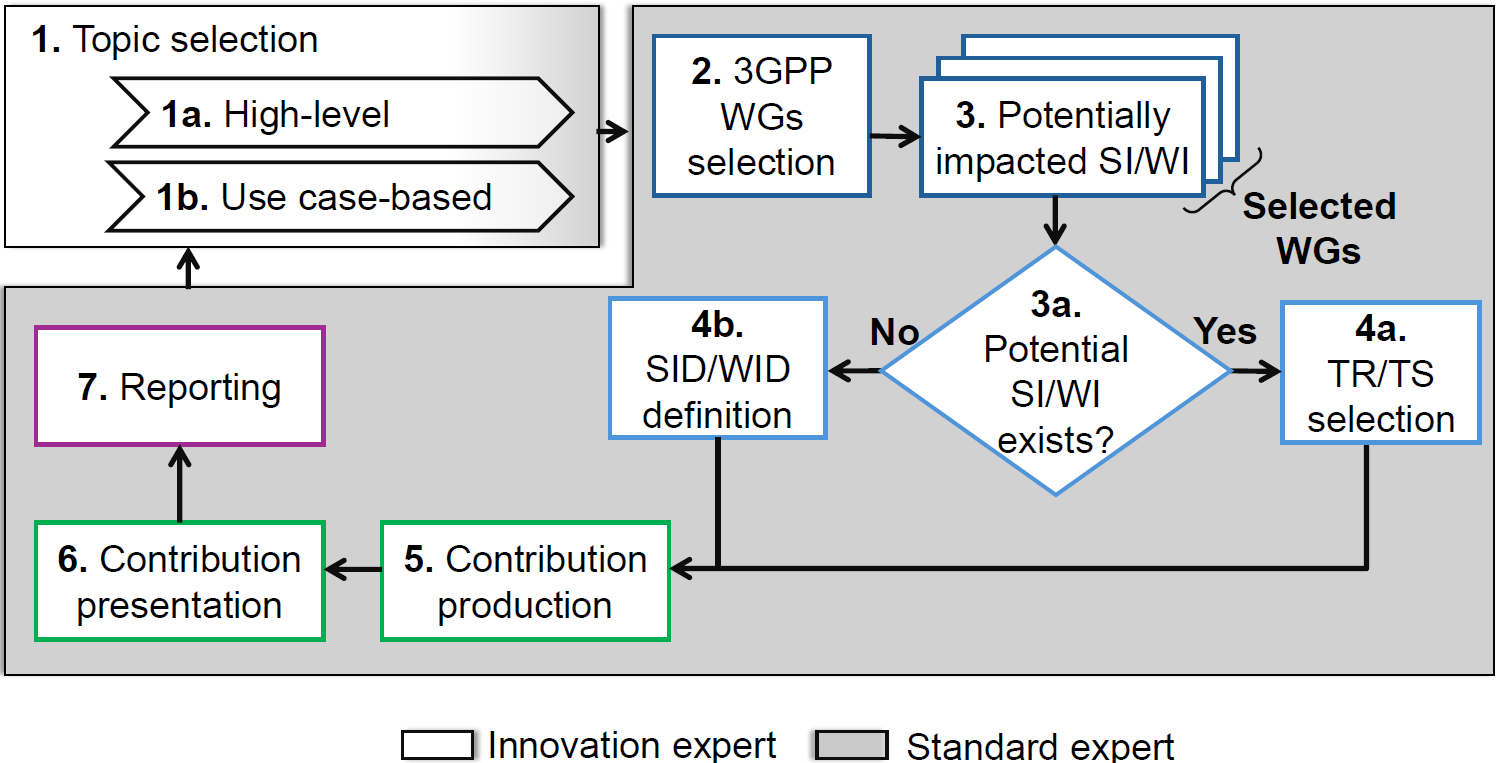}
\caption{Workflow of the proposed methodology for Standard contributions.}
\vspace{-0.3cm}
\label{fig_1}
\vspace{-0.3cm}
\end{figure}
However, bridging this gap remains challenging, as research outputs often lack structured pathways for integration into standardized frameworks. Without a systematic methodology that aligns research-driven innovations with standardization objectives, promising advancements risk remaining fragmented or underutilized. Addressing this issue requires a structured approach to evaluate, prioritize, and refine research contributions, ensuring alignment with strategic goals set by standardization bodies such as \gls{3gpp}. Strengthening collaboration between academia, industry, and regulatory organizations is essential to ensure that proposed solutions are both technically advanced and practically viable. Furthermore, establishing robust methodologies for transitioning research outcomes into standardization will be key to shaping an efficient, scalable, and future-proof 6G ecosystem.
This paper presents a structured methodology to bridge research and standardization in 6G development. The proposed multi-phase approach facilitates the systematic integration of research outcomes into standardized frameworks. A brief insight is then provided into its application, highlighting how research efforts from various RESTART projects on \glspl{mpn} and \glspl{ndt} can be translated into concrete standardization contributions. This ensures that research extends beyond academia, facilitating a seamless transition from innovation to global adoption.

The remainder of the paper is structured as follows: Sec.~\ref{sec:method} presents the proposed methodology for mapping research outcomes onto standardization frameworks. Sec.~\ref{sec:innovations} explores two key innovations studied within the RESTART initiative. Sec.~\ref{sec:analysis} illustrates the application of the proposed methodology to these innovations, while Sec.~\ref{sec:discussions} provides final considerations and future perspectives.

\section{Methodology for 3GPP Contribution}
\label{sec:method}
In this section, we present the key phases of the proposed methodology, designed to systematically bridge the gap between research and \gls{3gpp} contributions, ensuring effective translation of innovations into standardization efforts and real-world impact.

 % Originally developed to bring contributions from RESTART framework projects into standardization, this approach is inherently adaptable and generalizable to any research domain.

\subsection{Workflow Overview}
The proposed workflow, illustrated in Figure~\ref{fig_1}, unfolds across four interconnected phases, outlined below.

The \textbf{first phase} (step 1) focuses on the analysis of identified innovations and the collection of critical information to support the subsequent phases of the workflow. This phase involves gathering and organizing the following key information:
\begin{enumerate}
    \item general information about the innovation, including its context and relevance;
    \item a detailed description highlighting the innovation's technical characteristics and unique features;
    \item keywords associated with the innovation, aiding in categorization and targeting within standardization activities;
    \item system and scenario requirements that must be met for the innovation to be effectively applied;
    \item potential use cases demonstrating practical applications of the innovation.
    \item advantages and motivation justifying the innovation value and its alignment with broader objectives;
    \item references to related publications that provide a technical and scientific foundation for the innovation.
\end{enumerate}

\noindent This structured collection of information ensures that the innovation is thoroughly documented and strategically prepared for alignment with the objectives and processes of \gls{3gpp}. 
In particular, point 4 is crucial for identifying also possible limitations or constraints (e.g., industry adoption hurdles) that may affect the innovation’s path toward adoption and standardization.

% In the \textbf{second phase} (steps 2, 3 and 4), the relevant \gls{3gpp} \glspl{wg} are identified (step 2), along with the type of contributions to be produced (e.g. proposals for new Study or Work Items). This phase establishes a clear pathway for effectively addressing the technical and procedural requirements of the target WGs.

In the \textbf{second phase}, relevant \gls{3gpp} \glspl{wg} are identified (step 2), and for each, existing \glspl{wi} or \glspl{si} addressing the innovation topics are examined (Step 3). If these exist (step 3a), the associated \glspl{tr} and \glspl{ts} are selected for contribution (step 4a); otherwise, new \glspl{wi} or \glspl{si} are proposed (step 4b). This phase defines the pathway for addressing technical and procedural requirements of target \glspl{wg} and determining the type of contributions to be prepared (e.g., proposals for new \glspl{wi} or \glspl{si}).

The \textbf{third phase} (steps 5 and 6) is dedicated to the production and presentation of the technical contributions. This includes the preparation of comprehensive documentation that demonstrates the performance, feasibility, and applicability of the proposed innovations. These contributions are then actively presented and discussed during \gls{3gpp} \gls{wg} meetings to gather feedback and achieve consensus.

Finally, the \textbf{fourth phase} (step 7) involves documenting the outcomes of engagements, capturing key insights 
% and lessons 
learned from discussions and feedback within \gls{wg} meetings. These insights guide the development of subsequent contributions and inform necessary adjustments to ongoing research activities. This iterative approach ensures continuous improvement, fostering better alignment between research efforts and the evolving needs of the \gls{3gpp} standardization process.
\vspace{-0.1cm}
\subsection{Key Steps in the Contribution Process}

In the following, we provide a detailed breakdown of the key steps in the workflow outlined above.
\begin{itemize}
    \item \textbf{Mapping Research Innovations to 3GPP Needs}. The process starts with systematically mapping research innovations to the specific needs of \gls{3gpp}. This involves performing a gap analysis to identify how ongoing \gls{3gpp} standardization efforts align with research outcomes. This approach ensures that proposed innovations are both relevant and capable of addressing existing challenges in the standardization framework.
    \item \textbf{Developing New \glspl{sid} or \glspl{wid}}. Once relevant areas are identified, new \glspl{sid} or \gls{wid} are prepared to formally introduce the research topics to the relevant 3GPP working groups. These documents clearly articulate the problem, the objectives of the proposed work, and the expected outcomes.
    \item \textbf{Preparation of Technical Contributions}. Technical contributions are then developed, including detailed documentation of research findings, performance evaluations, simulation results, and use cases. These reports serve as the basis for discussions within 3GPP meetings. Visual aids, such as graphs, charts, and architectural diagrams, are also prepared to effectively communicate the technical aspects of the innovations.
    \item \textbf{Engaging in 3GPP Working Group Discussions}. Active participation in working group discussions is crucial. During these meetings, researchers present their findings, address questions, and incorporate feedback. This collaborative process helps refine the contributions and ensures that they meet the expectations of the working groups.
    \item \textbf{Building Consensus and Advocacy}. An essential component of this methodology is fostering collaboration and building consensus among \gls{3gpp} members. By addressing concerns and incorporating feedback, the research team can gain the necessary support for their proposals, increasing the likelihood of their adoption into the standardization process.
\end{itemize}

% By systematically aligning research outcomes with \gls{3gpp} processes, this methodology facilitates the transition of innovations from the lab to real-world implementation.

% This strategy is expected to not only facilitate the development
% of next-generation mobile networks but also to equip these
% networks to meet the diverse and evolving demands of the 6G
% era.

\section{Selected 6G Innovations in RESTART: Concepts and Applications}
\label{sec:innovations}
The proposed methodology has been applied to innovation topics explored in RESTART, specifically \glspl{mpn} and \gls{ndt}, to translate research outcomes into potential 6G standardization contributions. This section provides an overview of these two key innovations, highlighting their concepts, technical characteristics, and applications.
% With strong standardization potential, these innovations are pivotal to 6G advancement, embodying the forward-looking approaches essential for next-generation mobile networks.

% \vspace{-0.2cm}
\subsection{\glspl{mpn} for Cross-Layer - Cross-Domain Programmability}
\glspl{mpn} represent a transformative paradigm in network programmability, explicitly designed to address the challenges of 6G networks by introducing a new level of adaptability and intelligence~\cite{morphable_net}. \glspl{mpn} achieve this goal by enabling programmability both across protocol stack layers and different network domains, fostering a dynamic and responsive network environment. 
\glspl{mpn} are designed as an architecture where each network node can dynamically reconfigure its protocol stack (hence the name "Morphable"), supported by embedded \gls{ai}, \gls{ml}, and virtualization technologies. This enables real-time optimization and tailored support for a wide range of services, from \gls{xr} to autonomous transportation. \glspl{mpn} break traditional boundaries between layers and domains, creating a cohesive, end-to-end programmable infrastructure. This ensures that applications and devices achieve optimal performance and customization, regardless of underlying complexities and system dynamics.

In RAN3, the \gls{3gpp} \gls{wg} focusing on the overall UTRAN/E-UTRAN/NG-RAN architecture and protocols for the related network interfaces, efforts have centered on architectural enhancements for \glspl{mpn}, highlighting their ability to enable dynamic programmable network configurations across layers and domains. 

\subsubsection{Network programmability is key}
Programmable networks have emerged as a solution to overcome the lack of flexibility in traditional networks by allowing the customization of the behavior and functionalities within various protocol layers~\cite{survey_prog}. The key idea behind programmability is to embed custom code within various protocol layers and across technological domains, providing a foundation for flexible and efficient network operations. 

\glspl{mpn} go beyond vertical and horizontal programmability and leverage full programmability to achieve ultimate flexibility and robustness. Full programmability integrates vertical and horizontal approaches, creating a comprehensive framework for network-wide adaptability. This aims to integrate network-compute systems with end-to-end control, allowing the deployment of custom services across the programmable multidimensional continuum. This approach transforms the network into a highly adaptable platform to meet the specific needs of network owners, users, and applications.

%Network programmability can be categorized into two distinct approaches based on how programmability integrates within the network: vertical (cross-layer) and horizontal (cross-domain). Vertical programmability refers to programming functionality across different protocol stack layers, focusing on virtualizing and disaggregating network functions into flexible microservices. Horizontal programmability refers to programming functionalities across different network domains, enabling end-to-end integration and automation through standardized interfaces and protocols. 

\subsubsection{Implementation design}
\begin{figure}[t]
\centering
\includegraphics[width=0.9\columnwidth]{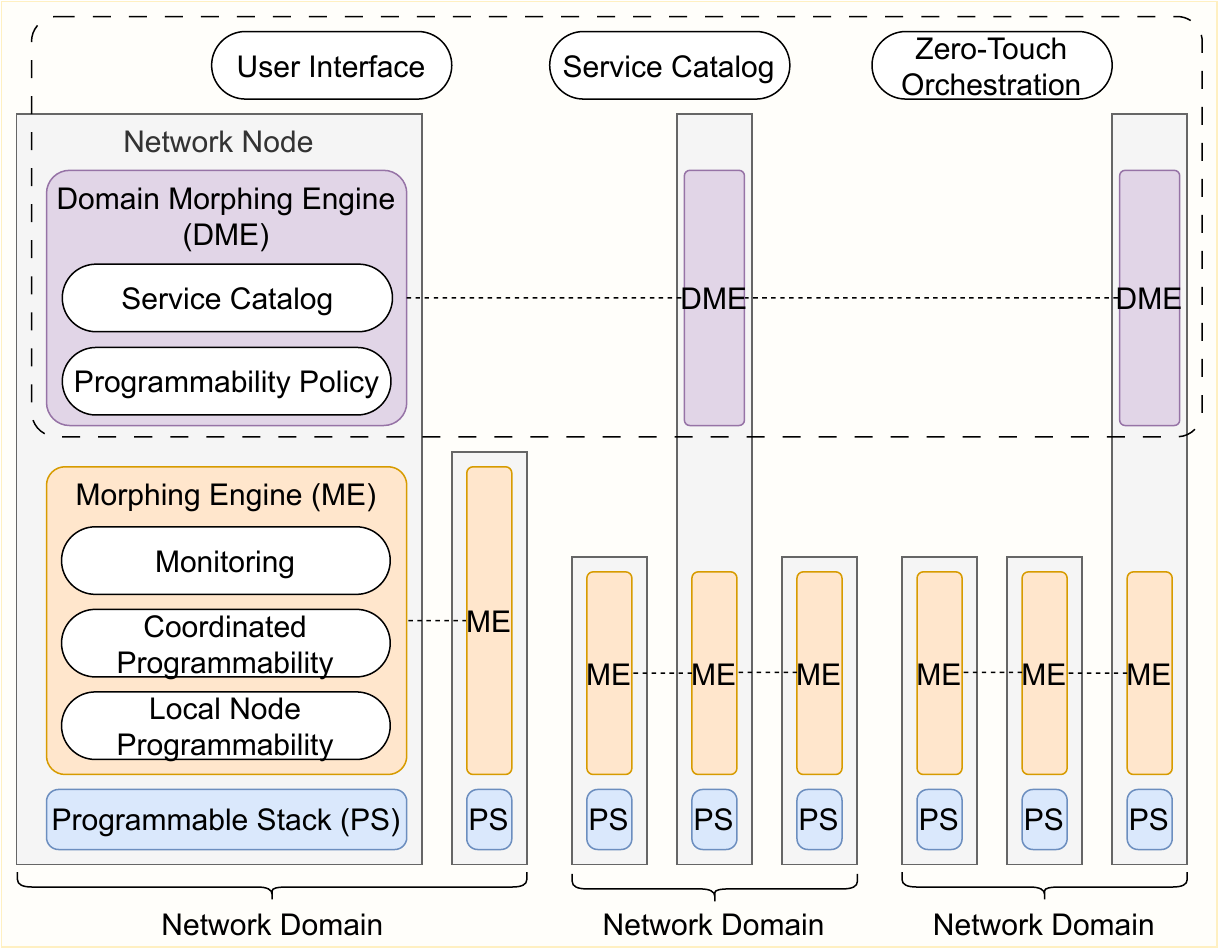}
\caption{MPN reference architecture, featuring MEs for each node and DMEs for each domain \cite{morphable_net}.}
\label{fig:mpn_arch}
% \vspace{-0.3cm}
\end{figure}

\glspl{mpn} build upon and extend existing virtualization and programmability techniques. Figure ~\ref{fig:mpn_arch} illustrates the \gls{mpn} reference architecture proposed in \cite{morphable_net}, where nodes can be reprogrammed from the physical to the application layer, hosting functions and protocols as containerized microservices or low-level code logic. This flexibility enables \glspl{mpn} to seamlessly adapt to diverse operational contexts and device heterogeneity.

At the core of \glspl{mpn} is a distributed yet coordinated management plane that optimizes the nodes protocol stack based on long-term network and environmental dynamics. The control and data planes enable real-time adaptation, ensuring dynamic network specialization in response to changing conditions.
%The management plane enables each MPN node to:
%\begin{itemize}
%\item Gather information on applications' and users' requirements, context, operational conditions, and other nodes' capabilities. 
%\item Select the best configuration of each protocol layer to ensure a consistent and efficient protocol stack within individual nodes and across multiple nodes and network domains. 
%\item Manage the lifecycle of deployed functions and protocols in coordination with other nodes.
%\end{itemize}
Each node incorporates a \gls{me} responsible for protocol layer adaptations and configurations, triggering adaptations of the intr-anode protocol stack by starting related microservices. \glspl{me} within a domain can elect a \gls{dme} to interact with similar entities in other domains. 
The data plane is envisioned to support real-time programmability, allowing dynamic adaptability to sudden changes in requirements and traffic. Run-time data plane programming goes beyond programming packet processing pipelines, offering broader customization possibilities through elements like host kernel stacks and smart network interface cards. 
% New programming models and abstractions, coupled with suitable Domain-Specific Languages (DSLs), are required to enable whole-network customization. 
The control plane ensures reliable service matching application requirements and configures the data plane consistently across multiple nodes. It reads network events from the data plane and preserves the nodes state during reprogramming. Network owners specify application and network requirements as high-level intents, which an intent compiler translates into control plane code. 
% \glspl{mpn} can exploit programmable hardware or specialized hardware to implement and adapt control protocols dynamically. 

\subsubsection{\glspl{mpn} management}
The \gls{E2E} management architecture of \glspl{mpn} spans different technological and administrative domains, creating a programmable network continuum. Management, or "morphing," is distributed within and across domains. As mentioned before, within each domain the ME at one node takes on the role of \gls{dme} through designation or election. The \gls{dme} interacts with other \glspl{dme} for network-wide service provisioning. 

\gls{E2E} network management is performed by scheduling the deployment and activation of functionalities at different layers of the Programmable Stack (PS). The \gls{dme} maintains a service catalog and establishes programmability policies to support coordination between \glspl{me}, ensuring consistent programmability within the domain. These policies may limit protocols and impose performance requirements, creating protocol stack blueprints for \glspl{me} to adapt based on application and service needs. 
Interactions between \glspl{dme} of different domains are crucial for cross-domain service provisioning, involving user interfaces, service databases, and zero-touch orchestration. Locally, each \gls{me} activates protocol stack blueprints based on local resource information and coordination with other \glspl{me}. The \gls{me} deploys functionalities at different protocol layers, optimizing them through dynamic compilation and optimization techniques to adapt to real-time network metrics. 

Services across domains are exposed by \glspl{dme} and can be requested by external users, including developers, administrators, external applications, and \glspl{dme} of other domains. Automatic operations within \glspl{mpn} are aided by \gls{ai}/\gls{ml}, which selects appropriate microservices, nodes, domains, protocols, and functionalities for deployment. 
% Different AI/ML approaches are used for service and system orchestration based on node characteristics, operational timescale, and data availability. 
The \gls{me} and \gls{dme} can be implemented as programmable components, drawing from O-RAN \cite{ORAN.WG1.OAM-Architecture} and \gls{3gpp} specifications \cite{3gpp.28.533}. 
%Each node can have a DME entity maintaining a service catalog and defining programmability policies, while an ME entity implements functions at different protocol layers. 
Open interfaces enable interactions among \glspl{me}, \glspl{dme}, protocol layers, and external users. Unlike O-RAN, \glspl{mpn} cover the entire protocol stack, support diverse node types, and dynamically adapt stack configurations to context, applications, and user needs. 
% \vspace{-0.3cm}
\subsection*{Use Case: MPNs for immersive XR applications}
%\textbf{Implementation of Morphable Networks.}
Immersive \gls{xr} applications serve as a compelling case study showcasing the benefits of \glspl{mpn}, as they require network performance beyond the limits of existing radio technologies~\cite{xr_challenges}. 

In a crowded stadium scenario, when a surge in \gls{xr} traffic occurs (e.g., during a pivotal game moment), \glspl{mpn} can detect the increased demand and immediately optimize the network across layers and domains. At the \gls{ran} level, \glspl{me} adjust \gls{mac}-layer parameters, allocating resources specifically for \gls{xr} traffic and refining scheduling and modulation schemes to maximize throughput. These adjustments are not isolated as they trigger coordinated changes in the transport layer at the edge domain, where protocols are fine-tuned to reduce latency and handle \gls{xr} interactive, low-latency data flows. Simultaneously, the cloud \gls{me} modifies the application-layer codecs, dynamically balancing video resolution and compression to accommodate varying connection qualities without compromising user experience. This real-time adaptability is guided by the \gls{dme}, which synchronizes the adjustments across the \gls{ran}, edge, and cloud domains. As user interactions shift, \glspl{mpn} reallocate resources and reconfigure protocols to ensure a seamless experience. Non-\gls{xr} traffic is deprioritized to maintain performance for \gls{xr} users, resulting in a network that evolves fluidly to meet changing demands.
% By integrating these real-time adaptations across layers and domains, \gls{mpn} maintain consistent quality of service and optimize resource utilization, ensuring that thousands of users can engage in immersive \gls{xr} experiences without disruption.

\begin{figure}[t]
\centering
\includegraphics[width=0.9\columnwidth]{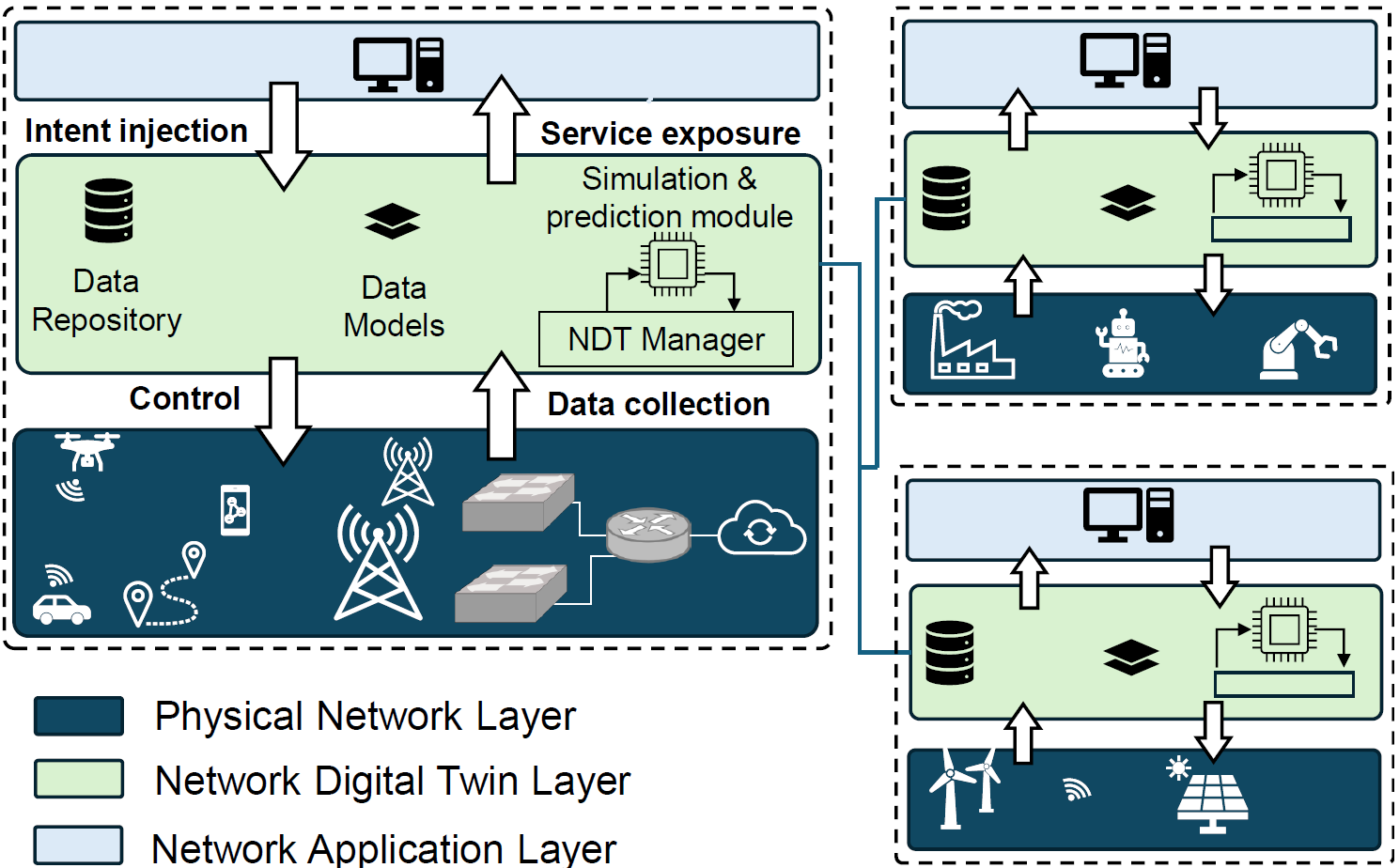}
\caption{Example architecture of a federated NDTs system integrating telecommunications, industrial, and electrical supply networks.}
\label{fig:ndt}
\end{figure}

%\vspace{0.2cm}
\subsection{\gls{ndt} for Mobile Networks}
The \gls{ndt} paradigm emerges as a transformative innovation in digital transformation, providing virtual models that precisely mirror real-world network configurations and dynamics.

In mobile networks, \gls{ndt} extends far beyond the conventional notion of \gls{dt}, which typically represents a digital replica of a physical, digital, or "phygital" asset \cite{DTN_24}. Instead, \glspl{ndt} form composite systems that integrate multiple interconnected components across the telecommunications infrastructure, serving as comprehensive virtual replicas of the entire network, encompassing all its elements, functions, and behaviors. In \gls{3gpp}, \gls{ndt} is addressed in SA5, where contributions have explored its integration for advanced network management through \glspl{ts} on Group Services, Management and Orchestration, and \gls{ndt} Management \cite{3gpp.28.915, 3gpp.28.561}. Similarly, \gls{itut} covers \gls{ndt} in its Recommendation on Digital Twin Network \cite{itut.y.3090}.

% By leveraging \gls{ndt} capabilities, telecommunications networks can achieve a higher degree of automation and operational efficiency, enhancing network management through \gls{ai}-driven automation and predictive maintenance. Real-time simulation enables proactive optimization and adaptive control, facilitating faster decision-making and more agile network reconfiguration. Moreover, NDT supports optimized resource allocation, improving network resilience, fault detection, and recovery mechanisms, ultimately leading to a more robust and self-adaptive network infrastructure. By minimizing trial-and-error deployments and enabling risk-free scenario testing, NDT significantly reduces operational costs and enhances overall efficiency, making it a highly valuable tool for telecom operators seeking to optimize network strategies in an increasingly complex and dynamic environment. 
% However, their realization and utilization rely on a set of key enabling aspects - some essential, while others enhance implementation, making their deployment more efficient, scalable, and effective. 

\subsubsection{Key enablers unlocking the power of NDTs}
The following outlines key enabling factors essential for unlocking the full potential of \glspl{ndt}. Some are fundamental, as their absence would make \gls{ndt} creation unfeasible, while others, though not strictly necessary, enhance efficiency, scalability, and overall effectiveness in both implementation and operational deployment. 
\begin{itemize}
\item{\textbf{Data Availability}.}
The realization of \glspl{ndt} relies on high-quality, multi-source data, categorized into geographic nationwide data and business-specific local data. The former includes territorial and network data, fostering transparency and cross-sector collaboration, while the latter comprises private, business-critical data for operational optimization and strategic decisions. Combined, they enable precise network modeling, long-term planning, and real-time decision-making.
\item{\textbf{\gls{sdn}}.}  
\gls{sdn} enables \glspl{ndt} by leveraging software controllers and \glspl{api} for efficient infrastructure management and seamless communication across modules, platforms, and networks. It optimizes data flow, enhances interoperability, and supports scalable microservices in cloud environments. Integrated with \gls{ml}, it drives predictive maintenance, cost efficiency, and performance optimization.

\item{\textbf{DevOps Software Development}.}  
\gls{devops} accelerates automation, collaboration, and iterative enhancements in \glspl{ndt}. Continuous Integration (CI) and Continuous Deployment (CD) ensure reliable updates and security compliance, while real-time data integration improves predictive capabilities. Open-source frameworks enhance interoperability, modularity, and scalability, enabling flexible \glspl{ndt} implementations.  

\item{\textbf{Management and Orchestration}.}  
Orchestration synchronizes data, integrates network domains, and optimizes resource management within \glspl{ndt}. Built on \gls{sba}, it enables dynamic interactions across platforms (e.g., operators, private networks, public administrations) and supports internal and cross-platform programmability. Leveraging cloud interoperability, zero-trust security, and open-source collaboration (e.g., ONAP, GSMA, O-RAN SC), \gls{sba}-driven orchestration enhances \glspl{ndt}, making them adaptive and scalable.
\end{itemize}

Finally, leveraging \gls{ml} algorithms and predictive analytics, \gls{ai} facilitates real-time analysis of vast datasets, enhancing the ability to anticipate future scenarios and optimize operations. It serves as the core engine driving strategic decisions and innovations in \glspl{ndt}, making it a foundational element across key enablers.

\subsubsection{NDT architectural overview}
Figure \ref{fig:ndt} illustrates an example architecture of a federated \glspl{ndt} system spanning telecommunications, industrial, and energy networks. Each of the dashed blocks in the figure follows a three-layer structure. The blue box represents the Physical Network Layer, which may span a single domain (e.g., access, transport, core, IP carrier) or an \gls{E2E} cross-domain system, encompassing all its elements and components.

At the core of the architecture, the \gls{ndt} Layer (green box) consists of three key subsystems:
i) Data Repository, which aggregates real-time network data collected from the Physical Network Layer for efficient storage, retrieval, and management, ensuring a highly accurate representation of real-world conditions.
ii) Data Models, which support \gls{ai}-based prediction, scheduling, and optimization, with basic models for network topology and state representation and functional models for analysis, emulation, and assurance.
iii) \gls{ndt} Entity Management, which oversees lifecycle, topology, model, and security management, ensuring integrity, visualization, and control of the \gls{ndt}.
The Network Application Layer (light-blue box) is responsible for network management, maintenance, and optimization. It interacts with the \gls{ndt} Layer through a structured exchange: the \gls{ndt} exposes its capabilities, offering a virtualized environment for analysis and simulation-based testing, while the Network Application Layer communicates its requirements via intent injection. The \gls{ndt} Layer processes these intents by modeling, simulating, and validating the requested functions before translating them into control updates for execution on the Physical Network, ensuring efficient and risk-minimized implementation.

\vspace{-0.1cm}
\subsection*{Use Case: federated NDTs for optimized and sustainable radio coverage}
If NDT already hold great potential, a federated NDT system further enhances its capability, enabling interconnected network digital twins to collaborate across multiple domains. This paradigm shift allows for a more comprehensive and dynamic representation of network operations, unlocking advanced optimization and cross-domain orchestration. 

In a smart city context, the municipal authority mandates specific radio planning requirements to ensure optimal coverage and compliance with urban policies. To meet these demands, a mobile network operator utilizes its radio coverage NDT and RIS NDT to optimize signal propagation and infrastructure deployment.
To enhance energy efficiency, the operator’s NDTs interface with the energy provider’s NDT, which models the availability of renewable energy resources. This collaboration allows the operator to dynamically adjust coverage and resource allocation based on real-time energy availability, ensuring a balance between network performance and sustainability.
By integrating these NDTs, the federated system enables multi-domain trade-offs, allowing the operator to assess whether increased signal strength is justified by energy consumption or if alternative strategies can optimize both coverage and efficiency.

\section{Selected Innovation Analysis}
\label{sec:analysis}

In this section, we illustrate how the proposed methodology has been applied to the two selected innovations explored within the RESTART framework.
The following presents the preliminary mapping phase, where the key concepts underlying \glspl{mpn} and \gls{ndt} have been correlated with thematic areas addressed within \gls{3gpp}. 
% This step is essential for positioning contributions within specific \glspl{wg} activities, facilitating the identification of the most relevant contexts for submitting standardization proposals.

% to facilitate the transition of innovations from the lab to real-world implementation.

% with a specific focus on analyzing the innovations that exhibit the strongest correlation with the use case.

% The potential success of these contributions highlights the value of a structured approach in bridging research and standardization.

 % The results provide a preliminary assessment of their key innovative aspects and their correlation with relevant technological topics addressed within the standardization framework.

% The objective is to provide an illustrative example of the methodology described earlier and its potential output. 
\vspace{-0.2cm}
\subsection{MNPs Analysis}
The concept of \glspl{mpn} introduces a paradigm shift in network programmability, designed to address the growing complexity and adaptability requirements of 6G networks.
The following aspects have been identified as strongly correlated with the \gls{mpn} paradigm:
\begin{itemize}
    \item \textbf{Network Architecture and Functionality}. \glspl{mpn} redefine network architecture by embedding programmability across multiple layers, enabling dynamic adaptability. The control, management, and data planes collectively support network-wide programmability, with the \gls{me} and \gls{dme} dynamically configuring network functionalities. This enhances protocol, interface, and resource flexibility, optimizing data transmission, traffic management, and service delivery through real-time programmability.
    \item \textbf{\gls{ai}/\gls{ml} for Network Optimization}. \gls{ai}/\gls{ml} in \glspl{mpn} enables intelligent protocol reconfiguration, dynamic service adaptation, and real-time traffic management. It supports micro-service selection, performance optimization, and automated decision-making based on network conditions, ensuring continuous adaptation to evolving demands.
\end{itemize}
The two aforementioned topics (Network Architecture and Functionalities, and \gls{ai}/\gls{ml} for Network Optimization) are closely aligned with ongoing discussions within \gls{3gpp} \glspl{wg}. Specifically, aspects related to network Architecture and Functionality are addressed within \gls{3gpp} SA2 and RAN3 \glspl{wg}, which focus on network design, interfaces, and functional splits within the \gls{5G} and emerging 6G ecosystems. Likewise, \gls{ai}/\gls{ml} for Network Optimization is a key focus of SA2, RAN3, and SA5, as these groups address \gls{ai}-driven network automation, resource optimization, and operational efficiency across both the core and \gls{ran} domains.
\vspace{-0.2cm}
\subsection{\gls{ndt} for Mobile Networks Analysis}
The \gls{ndt} for mobile networks is closely related to several key technological domains, each contributing to its effective implementation and operational efficiency. The following aspects have been identified as highly relevant to the \gls{ndt} paradigm:
\begin{itemize}
    \item \textbf{Digital Twin and Network Digital Twin}. As a virtual representation of a real-world network, \gls{ndt} aligns directly with the broader concept of \glspl{dt}, enabling real-time simulation, predictive maintenance, and adaptive control of mobile networks.
    \item \textbf{Resource Abstraction}. \gls{ndt} relies on resource abstraction to create a generalized data model that represents network and service-related resources independently of specific implementations, ensuring flexibility and interoperability.
    \item \textbf{AI/ML}. \gls{ai}-driven analytics play a crucial role in data analysis, predictive modeling, and intelligent automation within \glspl{ndt}, enabling real-time decision-making and self-adaptive network operations.
\end{itemize}
The identified topics (\gls{dt} and \gls{ndt}, Resource Abstraction, and \gls{ai}/\gls{ml}) are closely aligned with ongoing standardization efforts within \gls{3gpp}, particularly within SA5. SA5's work on \gls{ai}/\gls{ml}-driven network optimization and \gls{dt} frameworks plays a pivotal role in advancing \gls{ndt} solutions. Additionally, RAN3 and SA2 contribute primarily by defining \gls{ai}/\gls{ml}-driven mechanisms, ensuring seamless integration into network architecture and operations. Such a preliminary mapping must be complemented by an in-depth analysis of the ongoing activities within the identified \gls{3gpp} \glspl{wg}. i.e., step 3 and 4 in \ref{fig_1}. This additional investigation will refine the understanding of the current standardization landscape and support the development and submission of technical contributions within \gls{3gpp} (steps 5 and 6).

\section{Discussion and Future Perspectives}
\label{sec:discussions}
Bridging the gap between research and standardization is a complex challenge due to their differing priorities and methodologies. While research explores and validates innovative concepts, standardization focuses on ensuring interoperability, scalability, and real-world deployment feasibility. Aligning these perspectives is essential for translating cutting-edge technologies into standardized solutions.

This paper has proposed a structured methodology to facilitate this alignment, fostering collaboration between research community and standardization bodies. By integrating exploratory advancements into formalized frameworks, the approach ensures that innovations are both technically rigorous and industry-ready, supporting the development of scalable and interoperable 6G systems.

% As proof of the practical feasibility of this approach, a preliminary insight is provided to illustrate its application. 
This methodology has been applied to research conducted within RESTART projects on \glspl{mpn} and \glspl{ndt} \cite{3gpp_ndt_usecases_2025}, highly relevant topics in the evolution of next-generation networks. 
% \glspl{mpn} enable dynamic adaptability and efficient resource management, while \glspl{ndt} provide real-time simulation, predictive analytics, and intelligent decision-making. 
Their integration into 6G Release 20 will be pivotal in shaping a flexible and future-proof mobile ecosystem. 
The potential success of these contributions highlights the value of a structured methodology in bridging research and standardization, ensuring the seamless transition of innovations into industry-ready solutions. The underlying concepts and the overall workflow retain their validity, making the proposed methodology adaptable well beyond the 3GPP context. To maximize its effectiveness, it must evolve to support diverse research paradigms (from theoretical and system-level studies to experimental validation). 
Strengthening the research-standardization feedback loop will drive continuous innovation, accelerate technology adoption, and enable the development of robust, forward-looking standards that will define the next generation of mobile networks.

\section*{Acknowledgments}
This work was partially supported by the European Union - Next Generation EU under the Italian National Recovery and Resilience Plan (NRRP), Mission 4, Component 2, Investment 1.3, CUP F83C22001690001, partnership on “Telecommunications of the Future” (PE00000001 - program “RESTART”).

We would like to thank the contributors from the SUPER and IN projects, as well as from Grand Challenge 4, within the RESTART initiative for their valuable efforts in this work. 
\vspace{-0.4cm}
% \section{General Instructions}
% \begin{itemize}
%     \item Cite references: \cite{morphable_net}.

%     \item Cite figures: \ref{fig_1}.
%     \item Acronyms: \gls{ai}
% \end{itemize}

% {\appendix[Title]
% Text if needed.}
\small
\bibliographystyle{IEEEtran}
\bibliography{biblio}

\end{document}

%% file: acronyms.tex
\newacronym{2D}{2D}{2nd dimension}
\newacronym{3D}{3D}{3rd dimension}
\newacronym{3DN}{3DN}{3D Network}
\newacronym{3GPP}{3GPP}{3rd Generation Partnership Project}
\newacronym{4G}{4G}{Fourth-gGeneration}
\newacronym{5G}{5G}{Fifth-Generation}
\newacronym{6g}{6G}{Sixth-Generation}
\newacronym{wg}{WG}{Working Group}
\newacronym{3gpp}{3GPP}{3rd Generation Partnership Project}
\newacronym{sid}{SID}{Study Item Description}
\newacronym{wid}{WID}{Work Item Description}
\newacronym{ATG}{ATG}{Air-to-Ground}
\newacronym{BS}{BS}{Base Station}
\newacronym{CA}{CA}{Cooperative Awareness}
\newacronym{CAM}{CAM}{Cooperative Awareness Message}
\newacronym{CE}{CE}{Coverage Enhancement}
\newacronym{CP}{CP}{Collective Perception}
\newacronym{CPS}{CPS}{Collective Perception Service}
\newacronym{CPM}{CPM}{Collective Perception Message}
\newacronym{CSI}{CSI}{Channel State Information}
\newacronym{DRL}{DRL}{Deep Reinforcement Learning}
\newacronym{DDPG}{DDPG}{Deep deterministic policy gradient}
\newacronym{GoB}{GoB}{Grid-of-Beams}
\newacronym{GUE}{GUE}{Ground User Equipment}
\newacronym{ILP}{ILP}{Integer Linear Program}
\newacronym{IoT}{IoT}{Internet of Things}
\newacronym{ISD}{ISD}{inter-site distance}
\newacronym{ITS-S}{ITS-S}{Intelligent Transport System Station}
\newacronym{kpi}{KPI}{Key Performance Indicator}
\newacronym{LSP}{LSP}{large-scale parameter}
\newacronym{mMTC}{mMTC}{Massive Machine-type Communication}
\newacronym{mBS}{mBS}{Micro Base Station}
\newacronym{MBS}{MBS}{Macro Base Station}
\newacronym{LTE}{LTE}{Long Term Evolution}
\newacronym{LoS}{LoS}{Line of Sight}
\newacronym{NLoS}{NLoS}{Non-line of sight}
\newacronym{NB-IoT}{NB-IoT}{NarrowBand-IoT}
\newacronym{NFZ}{NFZ}{no-fly zone}
\newacronym{OFDM}{OFDM}{Orthogonal Frequency-Division Multiplexing}
\newacronym{O2O}{O2O}{Outdoor-to-Outdoor}
\newacronym{O2I}{O2I}{Outdoor-to-Indoor}
\newacronym{QoE}{QoE}{Quality of Experience}
\newacronym{QoS}{QoS}{Quality of Service}
\newacronym{RB}{RB}{Resource Block}
\newacronym{RL}{RL}{Reinforcement Learning}
\newacronym{RSSI}{RSSI}{Received Signal Strength Indicator}
\newacronym{RSRP}{RSRP}{Reference Signals Received Power}
\newacronym{RRM}{RRM}{Radio Resource Management}
\newacronym{RSU}{RSU}{Road Side Unit}
\newacronym{RU}{RU}{Resource Unit}
\newacronym{SAR}{SAR}{Search and Rescue}
\newacronym{SSB}{SSB}{Signal Synchronization Block}
\newacronym{SIC}{SIC}{successive interference cancellation}
\newacronym{SINR}{SINR}{Signal-to-Interference-plus-Noise ratio}
\newacronym{SNR}{SNR}{Signal-to-Noise ratio}
\newacronym{SSP}{SSP}{Small-scale Parameter}
\newacronym{TCP}{TCP}{Thomas Cluster Process}
\newacronym{TSP}{TSP}{Traveling Salesman Problem}
\newacronym{TR}{TR}{Technical Report}
\newacronym{UAS}{UAS}{Unmanned Aerial System}
\newacronym{UAV}{UAV}{Unmanned Aerial Vehicle}
\newacronym{UABS}{UABS}{Unmanned Aerial Base Station}
\newacronym{UE}{UE}{User Equipment}
\newacronym{UMa}{UMa}{Urban Macro}
\newacronym{UMi}{UMi}{Urban Micro}
\newacronym{UUE}{UUE}{Unmanned Aerial User Equipment}
\newacronym{UTM}{UTM}{UAS Traffic Management}
\newacronym{V2C}{V2C}{Vehicle-to-Cellular}
\newacronym{V2I}{V2I}{Vehicle-to-Infrastructure}
\newacronym{V2R}{V2R}{Vehicle-to-Roadside}
\newacronym{V2V}{V2V}{Vehicle-to-Vehicle}

\newacronym{RRA}{RRA}{Radio Resources Assignment}
\newacronym{PPP}{PPP}{Poisson Point Process}
\newacronym{pmf}{pmf}{Probability Mass Function}
\newacronym{CAV}{CAV}{Connected and Autonomous Vehicle}
\newacronym{mmWave}{mmWave}{Millimiter-Wave}
\newacronym{IRS}{IRS}{Intelligent Reflective Surface}
\newacronym{ITS}{ITS}{Intelligent Transport System}
\newacronym{E2E}{E2E}{End-to-End}
\newacronym{ULA}{ULA}{Uniform Linear Array}
\newacronym{RF}{RF}{Radio-Frequency}
\newacronym{HBF}{HBF}{Hybrid Digital Beamforming}
\newacronym{DFT}{DFT}{Discrete Fourier Transform}
\newacronym{sba}{SBA}{Service-Based Architecture}
\newacronym{itut}{ITU-T}{International Telecommunication Union - Telecommunication Standardization Sector}
\newacronym{sdo}{SDO}{Standard Development Organization}
\newacronym{devops}{DevOps}{Development and Operations}

\newacronym{b5g}{B5G}{Beyond 5G}
\newacronym{adc}{ADC}{Analog to Digital Converter}
\newacronym{aimd}{AIMD}{Additive Increase Multiplicative Decrease}
\newacronym{am}{AM}{Acknowledged Mode}
\newacronym{amc}{AMC}{Adaptive Modulation and Coding}
\newacronym{aqm}{AQM}{Active Queue Management}
\newacronym{balia}{BALIA}{Balanced Link Adaptation}
\newacronym{bdp}{BDP}{Bandwidth-Delay Product}
\newacronym{aqosa}{AQoSA}{Agile QoS Adaptation}
\newacronym{bf}{BF}{Beamforming}
\newacronym{cc}{CC}{Congestion Control}
\newacronym{cdf}{CDF}{Cumulative Distribution Function}
\newacronym{ccdf}{CCDF}{Complementary Cumulative Distribution Function}
\newacronym{NTN}{NTN}{Non-Terrestrial Network}
\newacronym{cn}{CN}{Core Network}
\newacronym{cqi}{CQI}{Channel Quality Information}
\newacronym{cp}{CP}{Control Plane}
\newacronym{csirs}{CSI-RS}{Channel State Information - Reference Signal}
\newacronym{dc}{DC}{Dual Connectivity}
\newacronym{dce}{DCE}{Direct Code Execution}
\newacronym{dmr}{DMR}{Deadline Miss Ratio}
\newacronym{dmrs}{DMRS}{DeModulation Reference Signal}
\newacronym{ecn}{ECN}{Explicit Congestion Notification}
\newacronym{edf}{EDF}{Earliest Deadline First}
\newacronym{enb}{eNB}{Evolved NodeB}
\newacronym{epc}{EPC}{Evolved Packet Core}
\newacronym{es}{ES}{Edge Server}
\newacronym{fdma}{FDMA}{Frequency Division Multiple Access}
\newacronym{fdd}{FDD}{Frequency Division Duplexing}
\newacronym[firstplural=Radio Access Technologies (RATs)]{rat}{RAT}{Radio Access Technology}
\newacronym{fs}{FS}{Fast Switching}
\newacronym{ftp}{FTP}{File Transfer Protocol}
\newacronym{gnn}{GNN}{Graph Neural Network}
\newacronym{gnb}{gNB}{Next Generation NodeB}
\newacronym{gbr}{GBR}{Guaranteed Bit Rate}
\newacronym{harq}{HARQ}{Hybrid Automatic Repeat request}
\newacronym{hetnet}{HetNet}{Heterogeneous Network}
\newacronym{hh}{HH}{Hard Handover}
\newacronym{hol}{HOL}{Head-of-Line}
\newacronym{ia}{IA}{Initial Access}
\newacronym{ieee}{IEEE}{Institute of Electrical and Electronics Engineers}
\newacronym{imt}{IMT}{International Mobile Telecommunication}
\newacronym{iot}{IoT}{Internet of Things}
\newacronym{ldpc}{LDPC}{Low-Density Parity Check}
\newacronym{los}{LOS}{Line-of-Sight}
\newacronym{lte}{LTE}{Long Term Evolution}
\newacronym{m2m}{M2M}{Machine to Machine}
\newacronym{ml}{ML}{Machine Learning}
\newacronym{mac}{MAC}{Medium Access Control}
\newacronym{mc}{MC}{Multi-Connectivity}
\newacronym{mcs}{MCS}{Modulation and Coding Scheme}
\newacronym{mec}{MEC}{Mobile Edge Cloud}
\newacronym{mi}{MI}{Mutual Information}
\newacronym{mimo}{MIMO}{Multiple Input, Multiple Output}
\newacronym{mptcp}{MPTCP}{Multipath TCP}
\newacronym{mr}{MR}{Maximum Rate}
\newacronym{mss}{MSS}{Maximum Segment Size}
\newacronym{mtd}{MTD}{Machine-Type Device}
\newacronym{mtu}{MTU}{Maximum Transmission Unit}
\newacronym{nfv}{NFV}{Network Function Virtualization}
\newacronym{nlosv}{NLOSv}{Vehicle Non-Line-of-Sight}
\newacronym{nr}{NR}{New Radio}
\newacronym{pdcch}{PDCCH}{Physical Downlink Control Channel}
\newacronym{pdcp}{PDCP}{Packet Data Convergence Protocol}
\newacronym{pdsch}{PDSCH}{Physical Downlink Shared Channel}
\newacronym{pdu}{PDU}{Packet Data Unit}
\newacronym{pf}{PF}{Proportional Fair}
\newacronym{ff}{FF}{Fairness First}
\newacronym{gi}{GI}{Guard Interval}
\newacronym{pdf}{PDF}{Probability Density Function}
\newacronym{pgw}{PGW}{Packet Gateway}
\newacronym{phy}{PHY}{Physical}
\newacronym{pbch}{PBCH}{Physical Broadcast Channel}
\newacronym[plural=\gls{mme}s,firstplural=Mobility Management Entities (MMEs)]{mme}{MME}{Mobility Management Entity}
\newacronym{pss}{PSS}{Primary Synchronization Signal}
\newacronym{pscch}{PSCCH}{Physical Sidelink Control Channel}
\newacronym{pucch}{PUCCH}{Physical Uplink Control Channel}
\newacronym{pusch}{PUSCH}{Physical Uplink Shared Channel}
\newacronym{rach}{RACH}{Random Access Channel}
\newacronym{ran}{RAN}{Radio Access Network}
\newacronym{red}{RED}{Random Early Detection}
\newacronym{rf}{RF}{Radio Frequency}
\newacronym{rlc}{RLC}{Radio Link Control}
\newacronym{rlf}{RLF}{Radio Link Failure}
\newacronym{rrc}{RRC}{Radio Resource Control}
\newacronym{rr}{RR}{Round Robin}
\newacronym{rs}{RS}{Remote Server}
\newacronym{rsrp}{RSRP}{Reference Signal Received Power}
\newacronym{rss}{RSS}{Received Signal Strength}
\newacronym{rtt}{RTT}{Round Trip Time}
\newacronym{rw}{RW}{Receive Window}
\newacronym{rx}{RX}{Receiver}
\newacronym{sa}{SA}{standalone}
\newacronym{sack}{SACK}{Selective Acknowledgment}
\newacronym{sap}{SAP}{Service Access Point}
\newacronym{sc}{SC}{Single Carrier}
\newacronym{sch}{SCH}{Secondary Cell Handover}
\newacronym{scoot}{SCOOT}{Split Cycle Offset Optimization Technique}
\newacronym{sdf}{SDF}{Service Data Flow}
\newacronym{sdma}{SDMA}{Spatial Division Multiple Access}
\newacronym{sinr}{SINR}{Signal to Interference plus Noise Ratio}
\newacronym{sl}{SL}{Supervised Learning}
\newacronym{sm}{SM}{Saturation Mode}
\newacronym{snr}{SNR}{Signal-to-Noise-Ratio}
\newacronym{son}{SON}{Self-Organizing Network}
\newacronym{ss}{SS}{Synchronization Signal}
\newacronym{srs}{SRS}{Sounding Reference Signal}
\newacronym{sss}{SSS}{Secondary Synchronization Signal}
\newacronym{tb}{TB}{Transport Block}
\newacronym{tcp}{TCP}{Transmission Control Protocol}
\newacronym{tdd}{TDD}{Time Division Duplexing}
\newacronym{TDMA}{TDMA}{Time Division Multiple Access}
\newacronym{tfl}{TfL}{Transport for London}
\newacronym{tm}{TM}{Transparent Mode}
\newacronym{trp}{TRP}{Transmitter Receiver Pair}
\newacronym{tti}{TTI}{Transmission Time Interval}
\newacronym{ttt}{TTT}{Time-to-Trigger}
\newacronym{tx}{TX}{Transmitter}
\newacronym{ue}{UE}{User Equipment}
\newacronym{UL}{UL}{Uplink}
\newacronym{uml}{UML}{Unified Modeling Language}
\newacronym{um}{UM}{Unacknowledged Mode}
\newacronym{utc}{UTC}{Urban Traffic Control}
\newacronym{vm}{VM}{Virtual Machine}
\newacronym{rsrq}{RSRQ}{Reference Signal Received Quality}
\newacronym{rssi}{RSSI}{Received Signal Strength Indicator}
\newacronym{crs}{CRS}{Cell Reference Signal}
\newacronym{nsa}{NSA}{Non Stand Alone}
\newacronym{mrdc}{MR-DC}{Multi \gls{rat} \gls{dc}}
\newacronym{endc}{EN-DC}{E-UTRAN-\gls{nr} \gls{dc}}
\newacronym{5gc}{5GC}{5G Core}
\newacronym{si}{SI}{Study Item}
\newacronym{wi}{WI}{Work Item}

\newacronym{iab}{IAB}{Integrated Access and Backhaul}
\newacronym{wf}{WF}{Wired-first}
\newacronym{hqf}{HQF}{Highest-quality-first}
\newacronym{pa}{PA}{Position-aware}
\newacronym{wbf}{WBF}{Wired Bias Function}
\newacronym{mib}{MIB}{Master Information Block}
\newacronym{sib}{SIB}{Secondary Information Block}
\newacronym{rnti}{RNTI}{Radio Network Temporary Identifier}
\newacronym{dft}{DFT}{Discrete Fourier Transform}
\newacronym{ppp}{PPP}{Poisson Point Process}
\newacronym{v2v}{V2V}{Vehicle-to-Vehicle}
\newacronym{wave}{WAVE}{Wireless Access in Vehicular Environments}
\newacronym{udp}{UDP}{User Datagram Protocol}
\newacronym{upa}{UPA}{Uniform Planar Array}
\newacronym{fec}{FEC}{Forward Error Correction}
\newacronym{V2X}{V2X}{Vehicle-To-Everything}
\newacronym{psfch}{PSFCH}{Physical Sidelink Feedback Channel}
\newacronym{pssch}{PSSCH}{Physical Sidelink Shared Channel}
\newacronym{csma}{CSMA}{Carrier Sense Multiple Access}
\newacronym{v2n}{V2N}{Vehicle-to-Network}
\newacronym{wlan}{WLAN}{Wireless Local Area Network}
\newacronym{v2i}{V2I}{Vehicle-to-Infrastructure}
\newacronym{d2d}{D2D}{Device-to-Device}
\newacronym{c-its}{C-ITS}{Connected Intelligent Transportation System}
\newacronym{fr2}{FR2}{Frequency Range 2}
\newacronym{fr1}{FR1}{Frequency Range 1}
\newacronym{sdu}{SDU}{Service Data Unit}
\newacronym{csi}{CSI}{Channel State Information}
\newacronym{scs}{SCS}{Subcarrier Spacing}
\newacronym{sumo}{SUMO}{Simulation of Urban MObility}
\newacronym{prr}{PRR}{Packet Reception Ratio}
\newacronym{edca}{EDCA}{Enhanced Distribution Channel Access}
\newacronym{sdap}{SDAP}{Service Data Adaptation Protocol}
\newacronym{iiot}{IIoT}{Industrial Internet of Things}
\newacronym{agv}{AGV}{Automated Guided Vehicles}
\newacronym{cm}{CM}{Configuration Management}
\newacronym{soa}{SoA}{State-of-the-Art}
\newacronym{snpn}{SNPN}{Standalone Non-Public Network}
\newacronym{pninpn}{PNI-NPN}{Public Network Interface Non-Public Network}
\newacronym{urllc}{URLLC}{Ultra-Reliable Low-Latency Communications}
\newacronym{qci}{QCI}{QoS Class Identifier}
\newacronym{ai}{AI}{Artificial Intelligence}
\newacronym{mab}{MAB}{Multi-Armed Bandit}
\newacronym{su}{SU}{Scheduling Unit}
\newacronym{ra}{RA}{Random Agent}
\newacronym{na}{NA}{Neural Agent}
\newacronym{ucba}{UCB-A}{UCB Agent}
\newacronym{tsa}{TS-A}{TS Agent}
\newacronym{ucb}{UCB}{Upper Confidence Bound}
\newacronym{ts}{TS}{Technical Specification}
\newacronym{nlts}{NLTS}{Neural Linear Thompson Sampling}
\newacronym{inf}{InF}{Indoor Factory}
\newacronym{me}{ME}{Morphing Engine}
\newacronym{dme}{DME}{Domain Morphing Engine}

\newacronym{infsl}{InF-SL}{Indoor Factory - Sparse Clutter, Low BS}
\newacronym{infdl}{InF-DL}{Indoor Factory - Dense Clutter, Low BS}
\newacronym{infsh}{InF-SH}{Indoor Factory - Sparse Clutter, High BS}
\newacronym{infdh}{InF-DH}{Indoor Factory - Dense Clutter, High BS}
\newacronym{us}{US}{Uplink Scheduler}
\newacronym{nn}{NN}{Neural Network}
\newacronym{dnn}{DNN}{Deep Neural Network}
\newacronym{das}{DAS}{Distributed Antenna System}
\newacronym{rl}{RL}{Reinforcement Learning}
\newacronym{embb}{eMBB}{Enhanced Mobile BroadBand}
\newacronym{5gacia}{5G-ACIA}{5G Alliance for Connected Industries and Automation}
\newacronym{cnn}{CNN}{Convolutional Neural Network}
\newacronym{aipm}{AIPM}{Artificial Intelligence Program Manager}
\newacronym{fci}{FCI}{Feedback Control Information}
\newacronym{prach}{PRACH}{Physical Random Access Channel}
\newacronym{coreset}{CORESET}{Control Resource Set}
\newacronym{cmab}{CMAB}{Contextual Multi-Armed Bandit}
\newacronym{drl}{DRL}{Deep Reinforcement Learning}
\newacronym{fd}{FD}{Frequency Domain}
\newacronym{td}{TD}{Time Domain}
\newacronym{mno}{MNO}{Mobile Network Operator}
\newacronym{bler}{BLER}{Block Error rate}
\newacronym{bl}{BL}{Bayesian Learning}
\newacronym{gml}{GML}{Machine Learning on Graphs}
\newacronym{bnn}{BNN}{Bayesian Neural Network}
\newacronym{pnn}{PNN}{Probabilistic Neural Network}
\newacronym{svi}{SVI}{Stochastic Variational Inference}
\newacronym{mape}{MAPE}{Mean Absolute Percentage Error}
\newacronym{ae}{AE}{Autoencoder}
\newacronym{vae}{VAE}{Variational Autoencoder}
\newacronym{lstm}{LSTM}{Long-Short Term Memory}
\newacronym{rnn}{RNN}{Recurrent Neural Network}
\newacronym{dci}{DCI}{Downlink Control Information}
\newacronym{rrf}{RRF}{Random Forest}
\newacronym{mlp}{MLP}{Multi-layer Perceptron}
\newacronym{rra}{MLP}{Multi-layer Perceptron}
\newacronym{mpn}{MPN}{Morphable Programmable Network}

\newacronym{api}{API}{Application Programming Interface}

\newacronym{hsdpa}{HSDPA}{High-Speed Downlink Packet Access}
\newacronym{xr}{XR}{Extended Reality}
\newacronym{dt}{DT}{Digital Twin}
\newacronym{ndt}{NDT}{Network Digital Twin}
\newacronym{ihrllc}{IHRLLC}{Immersive Hyper-Reliable Low Latency Communication}
\newacronym{bpsk}{BPSK}{Binary Phase Shift Keying}
\newacronym{qam}{QAM}{Quadrature Amplitude Modulation}
\newacronym{ber}{BER}{Bit Error Rate}
\newacronym{lau}{LAU}{Link Adaptation Unit}
\newacronym{amcs}{AMCS}{Adaptive Modulation and Coding Scheme}
\newacronym{adam}{ADAM}{Adaptive Moment Estimation}
\newacronym{c-rnti}{C-RNTI}{Cell Radio Network Temporary Identifier}
\newacronym{drx}{DRX}{Discontinuous Reception }
\newacronym{rop}{ROP}{Reporting Period}
\newacronym{pm}{PM}{Performance Measurement}
\newacronym{CM}{CM}{Configuration Management}
\newacronym{sla}{SLA}{Service Level Agreement}
\newacronym{sfs}{SFS}{Sequential Feature Selection}
\newacronym{oss}{OSS}{Operations Support System}
\newacronym{mlr}{MLR}{Multivariate Linear Regression} 
\newacronym{zsm}{ZSM}{Zero-touch Network and Service Management}
\newacronym{capex}{CAPEX}{Capital Expenditure}
\newacronym{opex}{OPEX}{Operational Expenditure}
\newacronym{awgn}{AWGN}{Additive White Gaussian Noise}
\newacronym{rmse}{RMSE}{Root Mean Square Error}
\newacronym{mdn}{MDN}{Mixture Density Networks}
\newacronym{k-nn}{K-NN}{K-Nearest Neighbors}
\newacronym{pqos}{PQoS}{Predictive Quality of Service}
\newacronym{fnn}{FNN}{Forward Neural Network}
\newacronym{volte}{VoLTE}{Voice over LTE}
\newacronym{dn}{DN}{Data Network}
\newacronym{nas}{NAS}{Non-Access Stratum}
\newacronym{5gs}{5GS}{5G Systems}
\newacronym{upf}{UPF}{User Plane Function}
\newacronym{up}{UP}{User Plane}

\newacronym{pdr}{PDR}{Packet Detection Rule}
\newacronym{qfi}{QFI}{QoS Flow Identifier}
\newacronym{pdb}{PDB}{Packet Delay Budget}
\newacronym{per}{PER}{Packet Error Rate}
\newacronym{arp}{ARP}{Allocation Retention Priority}
\newacronym{5qi}{5QI}{5G QoS Identifier}
\newacronym{drb}{DRB}{Data Radio Bearer}
\newacronym{an}{AN}{Access Network}

\newacronym{lr}{LR}{Learning rate}
\newacronym{voip}{VoIP}{Voice over Internet Protocol}
\newacronym{gmm}{GMM}{Gaussian Mixture Models}
\newacronym{gpd}{GPD}{Generalized Pareto Distributions }
\newacronym{cpu}{CPU}{Central Processing Unit}

\newacronym{DL}{DL}{Downlink}

\newacronym{tr}{TR}{Technical Report}

\newacronym{arq}{ARQ}{Automatic Repeat Request}

\newacronym{mmwave}{mmWave}{Millimeter-wave}

\newacronym{ns}{NS}{Network Slicing}
\newacronym{crc}{CRC}{Cyclic Redundancy Check}

\newacronym{knn}{KNN}{K-Nearest Neighbors}

\newacronym{prb}{PRB}{Physical Resource Block}

\newacronym{svr}{SVR}{Support Vector Regression}

\newacronym{fifo}{FIFO}{First-In, First-Out}
\newacronym{ip}{IP}{Internet Protocol}

\newacronym{sdn}{SDN}{Software Defined Networking}

\newacronym{v2x}{V2X}{Vehicle-to-Everything}

\newacronym{tbs}{TBS}{Transport Block Size}

\newacronym{ul}{UL}{Uplink}
\newacronym{ack}{ACK}{Acknowledge}

\newacronym{nack}{NACK}{Negative Acknowledgment}
\newacronym{saw}{SAW}{Stop-And-Wait}

\newacronym{kqi}{KQI}{Key Quality Indicator}

\newacronym{bs}{BS}{Base Station}

\newacronym{mmtc}{MMTC}{massive Machine Type Communications}

\newacronym{xai}{AI}{Explainable Artificial Intelligence}

\newacronym{ann}{ANN}{Artificial Neural Network}

\newacronym{fom}{FoM}{Figure of Merit}
\newacronym{rb}{RB}{Resource Block}

\newacronym{cu}{CU}{Centralized Unit}
\newacronym{du}{DU}{Distributed Unit}

\newacronym{nll}{NLL}{Negative log-likelihood}
\newacronym{kl}{KL}{Kullback-Leibler}

\newacronym{qoe}{QoE}{Quality of Experience}

\newacronym{bpnn}{BPNN}{Probabilistic Bayesian Neural Network}
\newacronym{shap}{SHAP}{Shapley Additive Explanations}
\newacronym{tfp}{TFP}{TensorFlow Probability}

\newacronym{mae}{MAE}{Mean Absolute Error}
\newacronym{p95th}{P95th}{95th Percentile}

\newacronym{u}{U}{User}